\documentclass[letterpaper, 10 pt, conference]{IEEEconf}
\usepackage{cite}
\usepackage{graphicx}
\usepackage[cmex10]{amsmath}
\usepackage{fixltx2e}
\usepackage{url}
\usepackage{widetext}
\usepackage{amsmath} 
 
\IEEEoverridecommandlockouts
\title{Bode's Sensitivity Integral Constraints: The Waterbed Effect Revisited}
\author{%
Abbas Emami-Naeini$^{\star}$ and
Dick~de~Roover$^\star$
\thanks{$^\star$A. Emami-Naeini and Dick de Roover are with
SC Solutions, Inc., 1261 Oakmead Pkwy, Sunnyvale, CA 94085, USA.}%
\thanks{$^{\ast}$Corresponding author: Abbas  Emami-Naeini: emami@scsolutions.com.}
}
\begin{document}
\maketitle  
\begin{abstract}
 Bode's sensitivity integral constraints define a fundamental rule about the limitations of feedback and is referred to as the waterbed effect. We take a fresh look at this problem and reveal an elegant and fundamental result that has been seemingly masked by previous derivations. The main result is that the sensitivity integral constraint is crucially related to the difference in speed of the closed-loop system as compared to that of the open-loop system. This makes much intuitive sense. Similar results are also derived for the complementary sensitivity function. In that case the integral constraint is related to the sum of the differences of the reciprocal of the transmission zeros and the closed-loop poles of the system. Hence all performance limitations are inherently related to the locations of the open-loop and closed-loop poles,  and the transmission zeros. A number of illustrative examples are presented.
\end{abstract}

\section{Introduction}
There is extensive literature on sensitivity of control systems and the fundamental and inevitable design limitations for linear time-invariant (LTI) systems [1]-[27]. One of the major contributions of Bode was to derive important fundamental and inescapable limitations on transfer functions that set limits on achievable design specifications. The majority of the previous results are based on Bode's sensitivity function,  $\mathcal{S}$, being the transfer function between the reference input to the tracking error or an output disturbance signal to the output (see Figure 1). Ideally we wish to have $|\mathcal{S}|\approx0$, which would provide perfect tracking and disturbance rejection. The sensitivity function is a measure of system sensitivity to plant variations [1]. In feedback control, the error in the overall transfer function gain is less sensitive to variations in the plant gain by a factor of $|\mathcal{S}|$ compared to errors in the open-loop control gain. For a unity feedback system as in Figure 1 with the loop gain $L(s)$, $n$ poles and $m$ finite transmission zeros, the reference input $r$, the output $y$, and the tracking error $e$,
\begin{equation}
E(s)=(I+L(s))^{-1}R(s)=\mathcal{S}(s)R(s),
\end{equation}%

\begin{figure}
\centering
\includegraphics[width=3in]{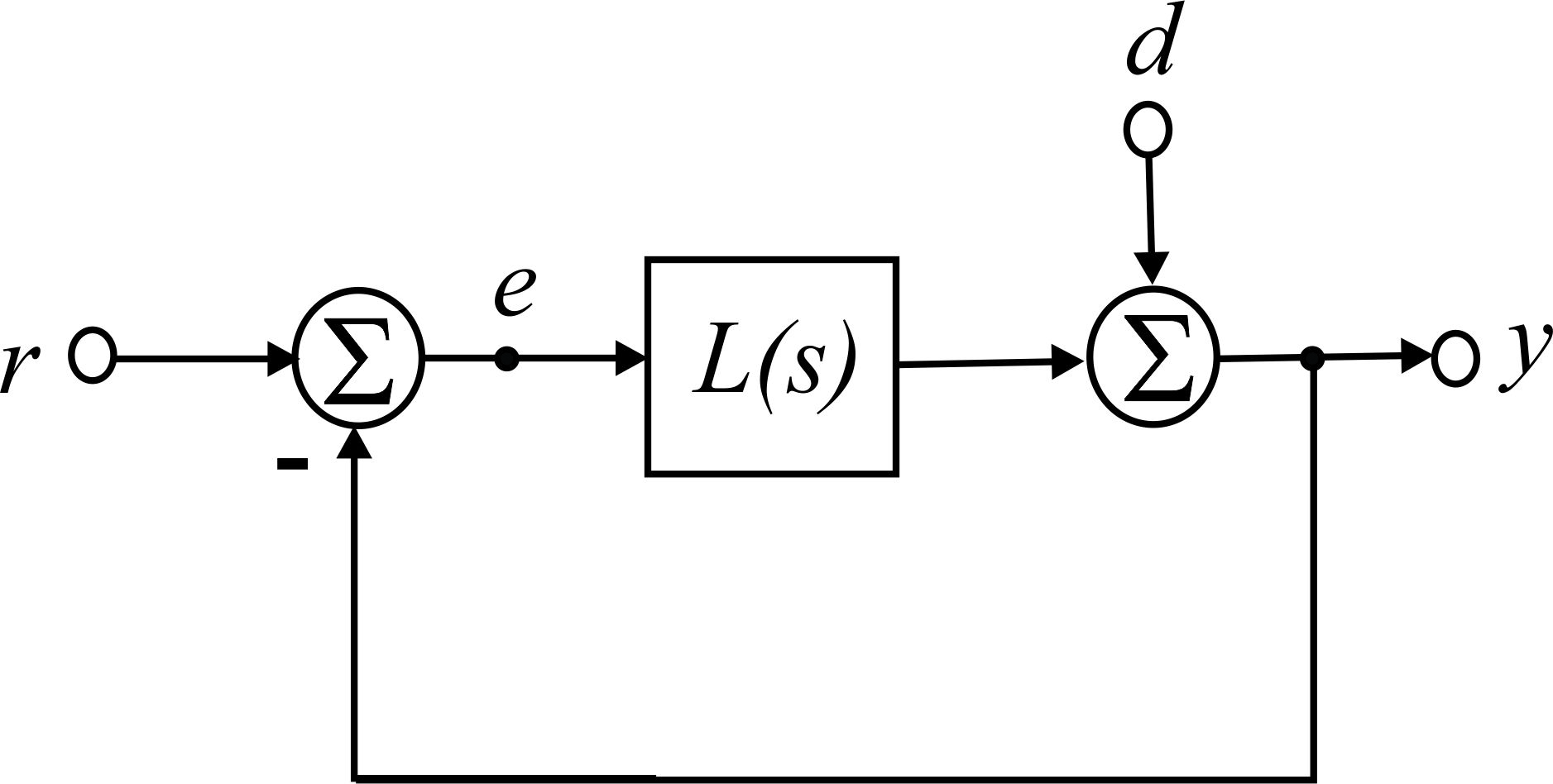}
\caption{Unity feedback system.}
\label{fig:fig1} 
\end{figure}
In addition to being the factor multiplying the system error, the sensitivity function, $\mathcal{S}$, is also the reciprocal of the distance of the Nyquist curve, $L(s)$, from the critical point (-1 point). A large $|\mathcal{S}_{max}|$ corresponds to a Nyquist plot that comes close to the $-1$ critical point and a system having a small complex margin [1], [29] that comes close to the point of instability. The frequency based specification based on the above equation can be expressed as
\begin{equation}
	|E|=|\mathcal{S}||R|<e_{b}. 
\end{equation}%
For minimum phase systems, the design rule was developed that the asymptotes of the Bode plot magnitude, which are restricted to be integral values for rational functions, should be made to cross over the zero-db line at a slope of $-1$ over a frequency range of about one decade around the crossover frequency [1]. An alternative to the standard Bode plot as a design guide can be based on a plot of the sensitivity function as a function of frequency. In this format, we require that the magnitude of the sensitivity function, $|\mathcal{S}|$, be less than a specified value $|\mathcal{S}|<1/W_1$) over the frequency range $0\leq\omega\leq\omega_1$ for tracking and disturbance rejection performance, and that $|\mathcal{S}|\approx1$ over the range $\omega_2 \leq\omega$ for stability robustness. Bode showed that for rational functions, with an excess of at least two more finite poles than zeros and no right half-plane (RHP) poles,
\begin{equation}
\int_{0}^{\infty}\ln(|\mathcal{S}|)d\omega=0.
\end{equation}%
 Eq. (3) represents a fundamental trade-off relationship in feedback control. It implies that if we make the log of the sensitivity function very negative (where $|\mathcal{S}|<1$) over some frequency band to reduce errors in that band, then, of necessity, $\ln(|\mathcal{S}|)$ will be positive (where $ |\mathcal{S}|>1$) over another part of the band, and errors will be amplified there (see Figure 2). Note that this figure is a log-linear plot (not log-log Bode plot). This means that the effect of disturbances are reduced for frequencies where $|\mathcal{S}|<1$ and they are amplified (an undesirable situation) for frequencies where $|\mathcal{S}|>1$. This characteristic is referred to as the "waterbed effect."  In Figure 2 we see that the area of disturbance attenuation is exactly balanced by the area of disturbance amplification as a result of Eq. (3).
\begin{figure}
\centering
\includegraphics[width=3in]{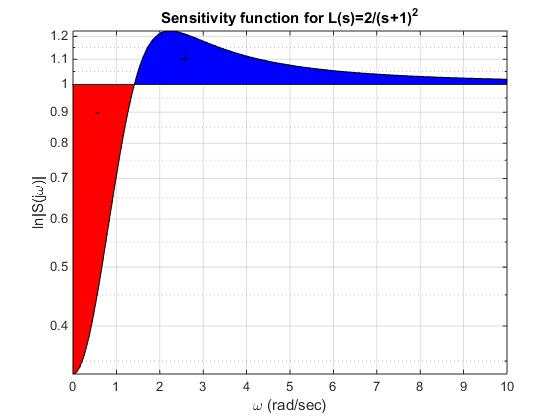}
\caption{Sensitivity function for second-order system: positive and negative areas cancel.}
\label{fig:fig2}
\end{figure}
    In addition, there is a fundamental algebraic constraint given by,
\begin{equation}
	\mathcal{S}+\mathcal{T}=I,
\end{equation}%
where the complementary sensitivity function is defined as
\begin{equation}
	\mathcal{T}(s)=(I+L(s))^{-1}L(s), 
\end{equation}%
which is the transfer function between the reference input $r$ and the output $y$ in Figure 1. Furthermore, at the RHP poles and zeros of the loop gain $L(s)$, we must satisfy the interpolation conditions [26]
\begin{equation}
	|\mathcal{T}(s)|_{s=p_{i}}=1,\quad   |\mathcal{T}(s)|_{s=z_{i}}=0.
\end{equation}%
\begin{equation}
	|\mathcal{S}(s)|_{s=z_{i}}=1,\quad   |{\mathcal{S}}(s)|_{s=p_{i}}=0.
\end{equation}%
    Gunter Stein suggests that we think of Bode's integral constraints as a kind of a conservation law and for the lack of any better terminology he refers to it as "conservation of sensitivity dirt," [25].  If performance improvements are sought in a frequency range (say low to midrange), then deterioration of performance must be tolerated in another frequency range (say high frequency range). In other words, "there is no free lunch!" If there are unstable poles, the situation is worse, because the positive area where the sensitivity magnifies the error must exceed the negative area where the error is reduced by the feedback. If the system is minimum phase, then it is, in principle, possible to keep the magnitude of the sensitivity small by spreading the sensitivity increase over all positive frequencies to infinity, but such a design requires an excessive bandwidth and is rarely practical. If a specific bandwidth is imposed, then the sensitivity function is constrained to take on a finite, possibly large, positive value at some point below the bandwidth resulting in large peak in the sensitivity function.
\subsection{Extensions of Bode Sensitivity Constraints}
 Bode's results have been extended for the open-loop unstable case. The constraint shows that the integral of the sensitivity function is determined by the presence of poles in the RHP. Suppose the loop gain $L(s)$ has $n_{p}$ poles, $\{p_{i}\}$, in the RHP and "rolls off" at high frequencies at a slope faster than $-1$. For rational functions, with an excess of at least two more finite poles than zeros, and $n_{p}$ unstable open-loop poles [10] showed that
\begin{equation}
	\int_{0}^{\infty}\ln(|\mathcal{S}|)d\omega=\pi\sum\limits_{i=1}^{n_{p}}\mathrm{Re}\{p_{i}\}.
\end{equation}%
If there are no RHP poles, then the integral is zero as before.
    If the system is not minimum-phase, the situation is even worse. An alternative to the above Eq. (8) is true if there is a nonminimum-phase zero of $L(s)$, a zero in the RHP. Suppose that the zero is located at $z_{o}=\sigma_{o}+j\omega_{o}$, where $\sigma_{o}>0$. Again, we assume there are $n_{p}$ RHP poles at locations $\{p_{i}\}$ with conjugate values $\{p_{i}^{*}\}$. Now the condition can be expressed as a two-sided weighted integral [10]
\begin{equation}
	\int_{-\infty}^{\infty}\ln(|\mathcal{S}|)W(z_0,\omega)d\omega=\pi\prod\limits\limits_{i=1}^{n_{p}}\ln\left\vert {\frac{p_{i}^{*}+z_{o}}{p_{i}-z_{o}}}\right\vert,
\end{equation}%
where
\begin{equation*}
W(z_0,\omega)=\frac{\sigma_{o}}{\sigma_{o}^{2}+(\omega-\omega_{o})^{2}}.
\end{equation*}
    In this case, we do not have the "roll-off" restriction, and there is no possibility of spreading the positive area over high frequencies, because the weighting function goes to zero with frequency. The important point about this integral is that if the nonminimum-phase zero is close to a RHP pole, the right side of the integral can be very large, and the excess of positive area is required to be correspondingly large. Based on this result, one expects especially great difficulty meeting both tracking and robustness specifications on the sensitivity with a system having RHP poles and zeros close together.
For the complementary sensitivity function [24] has shown that
\begin{equation}
\int\limits_{0}^{\infty }\frac{\ln \left\vert \mathcal{T}(j\omega
)\right\vert }{\omega ^{2}}d\omega =\pi
\sum\limits\limits_{i=1}^{i=n_{z}}\frac{1}{z_{i}}-\frac{\pi }{2K_{v}},
\end{equation}%
where
\begin{equation}
K_v=\lim_{s\rightarrow0}sL(s).
\end{equation}
 These results and their extensions have been the subject of intensive study and have provided great insight into the problem. We commend and strongly appreciate these previous studies [2]-[26]. However, we believe that the previous approaches were unknowingly hiding a fundamental result:\textit{ the sensitivity integral constraint is crucially related to the difference in speed (bandwidth) of the closed-loop system compared to that of the speed (bandwidth) of the open-loop system}.
    Our results reveal an elegant and fundamental relationship that has been seemingly masked by previous derivations. Hence the actual quantity is simply related to the shift in the poles of the system i.e. how fast the closed-loop poles of the system are compared to the open-loop poles. Furthermore, our derivation is direct and much simpler and does not rely on either Cauchy or Poisson-Jensen formulas that have been the focus of previous approaches to this problem.
    The organization of the rest of this Paper is as follows. In Section II we derive two fundamental relationships for the scalar case. One is a constraint on the sensitivity function and the other is a constraint on the complementary sensitivity function. Section III contains a SISO example. The same results are derived for the multivariable systems in Section IV. Section V provides two illustrative MIMO examples. Concluding remarks are in Section VI. The proofs of the theorems are contained in the Appendices.
\section{Sensitivity Constraints for SISO Systems}
In this section we present two theorems that establish constraints on the sensitivity and complementary sensitivity functions for single-input single-output (SISO) systems.\\
{\bf {Theorem 1}:} For any SISO closed-loop stable proper rational linear time-invariant (LTI) system Bode's integral
constraint may be described as%
\begin{equation}
\int\limits_{0}^{\infty }\ln \left\vert \mathcal{S}(j\omega )\right\vert
d\omega\mbox{}\\
=\left\{ 
\begin{array}{cl}
\frac{\pi }{2}\left( \sum\limits_{i=1}^{n}(\tilde{p}_{i}-p_{i})\right) , &\text{OLS} \\ 
\frac{\pi }{2}\left( \sum\limits_{i=1}^{n}(\tilde{p}_{i}-p_{i})+
\sum\limits_{i=1}^{n_{p}}2p_{i}\right) ,&  \text{OLU} \\
\end{array} 
\right. 
\end{equation}%
where $\mathcal{S}$ is the sensitivity function, $\{\tilde{p}%
_{i}\}, i=1,2,\ldots,n$ are the locations of the closed-loop poles and, $%
\{p_{i}\}$, $i=1,2,\ldots,n$ are the locations of the open-loop poles, and there
are possibly $n_{p}$ unstable open-loop poles at $\{p_{i}\}, i=1,...,n_{p}$
(including multiplicities) with $\textrm{Re}\{p_{i}\}>0$. OLS refers to an open-loop stable system and OLU refers to an open-loop unstable system.\\
\begin{proof}
See Appendix A. The proof is similar to the one in [10].
\end{proof} 
The fundamental relationship is that the sum of the areas underneath the $
\ln(\left\vert \mathcal{S}\right\vert )$ curve is related to the difference in
speeds of the closed-loop and open-loop systems. This makes a lot of sense
from an intuitive point of view. If the system is open-loop unstable then
additional positive area is added leading to further sensitivity
deterioration.\\
{Corollary 1:} Previous results [24] have shown that if $L(s)$ is strictly proper then
\begin{equation}
\int\limits_{0}^{\infty }\ln \left\vert \mathcal{S}(j\omega )\right\vert
d\omega\mbox{}\\
=\pi\sum\limits_{i=1}^{n_p}p_{i}-\frac{\pi K_h}{2},\\
\end{equation}%
\begin{equation}
K_h=\lim_{s\rightarrow \infty} sL(s).
\end{equation}
It can be readily shown that for a strictly proper system
\begin{equation}
K_h=-\sum\limits_{i=1}^{n}{\tilde p}_{i}+\sum\limits_{i=1}^{n}p_{i},
\end{equation}
which when substituted in Eq. (13) yileds the same result as in Eq. (12). The following theorem refers to the concept of system Type [1].\\
{\bf{Theorem 2}:} For any SISO closed-loop stable proper rational LTI system the complementary sensitivity integral constraint may be described by
\begin{equation}
\begin{split}
&\int\limits_{0}^{\infty }\frac{\ln \left\vert \mathcal{T}(j\omega
)\right\vert }{\omega ^{2}}d\omega =\\
&\left\{  
\begin{array}{cl}
\infty,  & \text{Type 0 system} \\ 
\frac{\pi }{2}\left( \sum\limits_{i=1}^{i=n}\frac{%
1}{\tilde{p}_{i}}-\sum\limits_{i=1}^{i=m}\frac{1}{z_{i}}\right) +\pi \sum\limits_{i=1}^{i=n_{z}}\frac{1}{z_{i}}, &\text{%
Type I (or higher)}%
\end{array}%
\right. 
\end{split}
\end{equation}
where $\{-\tilde{p}_{i}\}$\ are the closed-loop poles, $\{-z_{i}\}, i=1,\ldots,m$ are
the closed-loop transmission zeros and there are possibly $n_z$ non-minimum
phase transmission zeros of the system (including multiplicities) with $\textrm{Re}\{-z_{i}\}>0$.\\
\begin{proof}
See Appendix B.
\end{proof}
{Corollary 2:}
Using Truxal's identity [3], [1] for a Type I system we have
that,%
\begin{equation}
\int\limits_{0}^{\infty }\frac{\ln \left\vert \mathcal{T}(j\omega
)\right\vert }{\omega ^{2}}d\omega =-\frac{\pi }{2K_{v}}+\pi
\sum\limits\limits_{i=1}^{i=n_{z}}\frac{1}{z_{i}}.
\end{equation}%
where $K_{v}$ is the velocity error coefficient [1], and this result is in agreement with previous results Eq. (10).\\
{Corollary 3:}
Since complex conjugate poles come in pairs and the imaginary parts always
cancel each other, the result (for computation) simplifies to%

\begin{equation}
\begin{split}
&\int\limits_{0}^{\infty }\ln \left\vert \mathcal{S}(j\omega )\right\vert
d\omega \\
&=\begin{cases}
\frac{\pi }{2}\left(\sum\limits_{i=1}^{n}(\textrm{Re}\{\tilde{p}_{i}\}-\textrm{Re}\{p_{i}\}) \right), & \text{OLS} \\ 
\frac{\pi }{2}\left(\sum\limits_{i=1}^{n}(\textrm{Re}\{\tilde{p}_{i}\}-\textrm{Re}
\{p_{i}\})+\sum\limits_{i=1}^{n_{p}}2\textrm{Re}\{p_{i}\}\right), & \text{OLU} 
\end{cases}
\end{split}
\end{equation}

\section{SISO Example}
We present a SISO example to illustrate the results.
{Example 1}:
Consider the system with the loop gain%
\begin{equation*}
L(s)=\frac{(s+1)}{s^{2}}.
\end{equation*}%
The sensitivity function is
\begin{equation*}
\mathcal{S}(s)=(I+L)^{-1}=\frac{s^{_{2}}}{s^{2}+s+1},
\end{equation*}%
\begin{equation*}
\left\vert \mathcal{S}(j\omega )\right\vert =\frac{\omega ^{2}}{\sqrt{%
(-\omega ^{2}+1)^{2}+\omega ^{2}}},
\end{equation*}%
\begin{equation*}
\int\limits_{0}^{\infty }\ln \left\vert \mathcal{S}(j\omega )\right\vert
d\omega =-\frac{\pi }{2},
\end{equation*}%
which agrees with the answer from Eq. (12),
\begin{equation*}
\int\limits_{0}^{\infty }\ln \left\vert \mathcal{S}(j\omega )\right\vert
d\omega =\frac{\pi }{2}(-0.5-0-0.5-0)=-\frac{\pi }{2}.
\end{equation*}%
\begin{figure}
\centering
\includegraphics[width=3in]{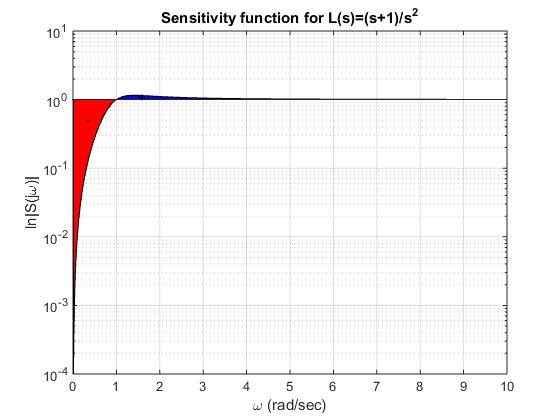}
\caption{Sensitivity function for Example 1.}
\label{fig:fig4}
\end{figure}
A plot of the log magnitude of the sensitivity function is shown in Figure 4. The negative area exceeds the positive area by $-\frac{\pi }{2}$%
. Hence more improvement in sensitivity is possible for the frequency range
below $\omega =1$ rad/sec as compared to the amount of deterioration for $%
\omega >1$ rad/sec. The complementary sensitivity function is 
\begin{equation*}
\mathcal{T}(s)=(I+L(s))^{-1}L(s)=\frac{s+1}{s^{2}+s+1},
\end{equation*}%
\begin{equation*}
\left\vert \mathcal{T}(j\omega )\right\vert =\frac{\sqrt{\omega ^{2}+1}}{%
\sqrt{(-\omega ^{2}+1)^{2}+\omega ^{2}}}.
\end{equation*}%
Since the system is Type II,
\begin{equation*}
\int\limits_{0}^{\infty }\frac{\ln \left\vert \mathcal{T}(j\omega
)\right\vert }{\omega ^{2}}d\omega =0.
\end{equation*}%
Using our formula, Eq. (16), we obtain the same answer
\begin{equation*}
\begin{split}
\int\limits_{0}^{\infty }\frac{\ln \left\vert \mathcal{T}(j\omega
)\right\vert }{\omega ^{2}}d\omega
&=\frac{\pi }{%
2}\left( \frac{1}{1}-\left( \frac{1}{\frac{1}{2}+j\frac{\sqrt{3}}{2}}+\frac{1}{%
\frac{1}{2}-j\frac{\sqrt{3}}{2}}\right) \right) =0.
\end{split}
\end{equation*}%
\section{Sensitivity Constraints for Multivariable Systems}
In this section we present two theorems that establish constraints on the sensitivity and complementary sensitivity functions for MIMO systems.\\
{{\bf Theorem 3}:} For any square (and non-singular) LTI MIMO system with no hidden modes,
Bode's sensitivity integral constraint may be described as%
\begin{equation}
\int\limits_{0}^{\infty }\ln \left\vert \det \left[ \mathcal{S}(j\omega )%
\right] \right\vert d\omega =\left\{ 
\begin{array}{cl}
\frac{\pi }{2}\left( \sum\limits\limits_{i=1}^{n}(\tilde{p}_{i}-p_{i})\right), \quad\text{%
OLS} \\ 
\frac{\pi }{2}\left( \sum\limits\limits_{i=1}^{n}(\tilde{p}_{i}-p_{i})+%
\sum\limits\limits_{i=1}^{n_{p}}2p_{i}\right), \text{OLU}%
\end{array}%
\right.
\end{equation}%
where $\mathcal{S}$ is the sensitivity function, $\{\tilde{p}_{i}\}, i=1,..., n$, are
the locations of the closed-loop poles and, $\{p_{i}\},  i=1,..., n$, are the locations
of the open-loop poles, and there are possibly $n_{p}$ unstable open-loop poles
(including multiplicities) with $\mathrm{Re}\{p_{i}\}>0, i=1,\ldots,n_{p}$. OLS refers to an open-loop stable system and OLU refers to an open-loop unstable system.\\
\begin{proof}
See Appendix C.
\end{proof}
The authors believe that this is the first direct derivation of this result for the MIMO case. It is seen that unlike the results in [20, page 146] and [24, page 782] the constraint on sensitivity, in this unweighted form,
is not dependent on the directions of the poles and their relative
interaction. The following theorem refers to the concept of system Type for multivariable systems [28].\\
{{\bf Theorem 4}:} For any closed-loop stable square (and non-singular) Type [1 1\ldots1] MIMO system, with no hidden
modes, the integral constraint on the complementary sensitivity function may be
described as
\begin{equation}
\begin{split}
&\int\limits_{0}^{\infty }\frac{1}{\omega ^{2}}\ln \left\vert \det \left[ 
\mathcal{T}(j\omega )\right] \right\vert d\omega \\
&= \left\{ 
\begin{array}{cc}
\infty, & \text{Type [0 0\ldots0]} \\ 
\frac{\pi }{2}\left(\sum\limits\limits_{i=1}^{i=n}\frac{%
1}{\tilde{p}_{i}}-\sum\limits\limits_{i=1}^{i=m}\frac{1}{z_{i}}\right) +\pi \sum\limits\limits_{i=1}^{i=n_{z}}\frac{1}{z_{i}}, & \text{%
Type [1 1\ldots1]}%
\end{array}%
\right.
\end{split}
\end{equation}%
where $\{\tilde{p}_{i}\},  i=1,..., n$\ are the closed-loop poles, $\{z_{i}\}, i=1,\ldots, m$ (including multiplicities) are
the (finite) closed-loop transmission zeros and $\{z_{i}\}, i=1\ldots,n_{z}$ are the non-minimum
phase transmission zeros of the system.\\
\begin{proof}
See Appendix D.
\end{proof}
The authors believe that this is the first direct derivation of this result for the MIMO case. Again this shows that the constraint on the complementary sensitivity
function is not dependent on the pole/zero directions either. We now
demonstrate the merits of the above results with two multivariable
examples.
\section{MIMO Examples}
{Example 2}: Consider the system [30] with the loop gain
\begin{equation*}
L(s)=\frac{1}{(s+1)(s+2)}\left[ 
\begin{array}{cc}
-47s+2 & 56s \\ 
-42s & 50s+2%
\end{array}%
\right] ,
\end{equation*}%
that has open-loop poles at $-1$,$-2$, and a no finite transmission zeros.
The sensitivity function is 
\begin{eqnarray*}
\begin{split}
\mathcal{S}(s) &=(I+L)^{-1}\\
&=\frac{1}{(s+2)(s+4)}\left[ 
\begin{array}{cc}
s^{2}+53s+4 & -56s \\ 
42s & s^{2}-44s+4
\end{array}%
\right] , \\
\end{split}
\end{eqnarray*}
\begin{eqnarray*}
\det [\mathcal{S}(s)] &=&\frac{(s+1)(s+2)}{(s+2)(s+4)},\\
\end{eqnarray*}%
\begin{equation*}
\left\vert \det [\mathcal{S}(j\omega )\right]\vert =\frac{\sqrt{(\omega
^{2}+1)(\omega ^{2}+4)}}{\sqrt{(\omega ^{2}+4)(\omega ^{2}+16)}}.
\end{equation*}%
\begin{equation*}
\int\limits_{0}^{\infty }\ln \left\vert \det \left[ \mathcal{S}(j\omega )%
\right] \right\vert d\omega =-\frac{3\pi }{2}.
\end{equation*}%
Using our formula, Eq. (19), we find the same answer
\begin{equation*}
\int\limits_{0}^{\infty }\ln \left\vert \det \left[ \mathcal{S}(j\omega )%
\right] \right\vert d\omega =\frac{\pi }{2}(-2-4+1+2)=-\frac{3\pi }{2}.
\end{equation*}
 The complementary sensitivity function is%
\begin{equation*}
\begin{split}
\mathcal{T}(s)&=(I+L)^{-1}L\\
&=\frac{1}{(s+2)(s+4)}\left[ 
\begin{array}{cc}
-47s+4 & 56s \\ 
-42s & 50s+4%
\end{array}%
\right] .
\end{split}
\end{equation*}%
\begin{equation*}
\det \left[ \mathcal{T}(s)\right] =\frac{2}{(s+2)(s+4)}.
\end{equation*}%
\begin{equation*}
\left\vert \det \left[ \mathcal{T}(j\omega )\right] \right\vert =\frac{2}{%
\sqrt{(\omega ^{2}+4)(\omega ^{2}+16)}}.
\end{equation*}%
This is a Type [0 0] system and therefore from Eq. (20),
\begin{equation*}
\int\limits_{0}^{\infty }\frac{\ln \left\vert \det \left[ \mathcal{T}%
(j\omega )\right] \right\vert }{\omega ^{2}}d\omega =\infty .
\end{equation*}
{Example 3 [32, page 174]}:
Consider the Type [1 1] system with the loop gain
\begin{equation*}
\begin{split}
L(s)&=\left[ 
\begin{array}{cc}
\frac{1}{s(s+1)} & \frac{1}{s+1} \\ 
\frac{1}{s} & \frac{4}{s(s+2)}%
\end{array}%
\right] \\
&=\left[ 
\begin{array}{cc}
s(s+1) & 0 \\ 
0 & s(s+2)%
\end{array}%
\right] ^{-1}\left[ 
\begin{array}{cc}
1 & s \\ 
s+2 & 4%
\end{array}%
\right],
\end{split}
\end{equation*}%
that has open-loop poles at $0$, $0$, $-1$, $-2$, and finite transmission zeros
at $+1.2361$ and $-3.2361$. The sensitivity function is 
\begin{eqnarray*}
\begin{split}
\mathcal{S}(s) &=(I+L)^{-1}\\&=\frac{s^{2}(s+1)(s+2)}{s^{4}+3s^{3}+6s^{2}+4s+4}%
\end{split}
\left[ 
\begin{array}{cc}
\frac{(s+2)^{2}}{s(s+2)} & -\frac{1}{s+1} \\ 
-\frac{1}{s} & \frac{s^{2}+s+1}{s(s+1)}%
\end{array}%
\right] , \\
\begin{split}
\det[\mathcal{S}(s)]=\frac{s^{2}(s+1)(s+2)}{s^{4}+3s^{3}+6s^{2}+4s+4}.
\end{split}
\end{eqnarray*}%
\begin{equation*}
\left\vert \det \left[ \mathcal{S}(j\omega )\right] \right\vert =\left\vert 
\frac{-\omega ^{2}\sqrt{(\omega ^{2}+1)(\omega ^{2}+4)}}{\sqrt{(\omega
^{4}-6\omega ^{2}+4)^{2}+(-3\omega ^{3}+4\omega )^{2}}}\right\vert
\end{equation*}%
\begin{equation*}
\int\limits_{0}^{\infty }\ln \left\vert \det \left[ \mathcal{S}(j\omega )%
\right] \right\vert d\omega =0.
\end{equation*}%
Using our formula, Eq. (19), we have that
\begin{equation*} 
\begin{split}
&\int\limits_{0}^{\infty }\ln\left\vert  \det \left[ \mathcal{S}(j\omega )
\right] \right\vert d\omega=\frac{\pi }{2}(-1.3277\\
&-1.3277-0.1723-0.1723+1+2+0+0)=0. \\
\end{split}
\end{equation*}%
The complementary sensitivity
function is 
\begin{eqnarray*}
\mathcal{T}(s) &=&(I+L(s))^{-1}L(s), \\
&=&\frac{1}{\Delta(s)}\left[ 
\begin{array}{cc}
4 & s^{2}(s+2) \\ 
s(s+1)(s+2) & 3s^{2}+2s+4%
\end{array}%
\right] , \\
\det [\mathcal{T}(s)] &=&\frac{-s^{2}-2s+4}{s^{4}+3s^{3}+6s^{2}+4s+4}.
\end{eqnarray*}%
where $\Delta(s)=s^{4}+3s^{3}+6s^{2}+4s+4$.
\begin{equation*}
\left\vert \det [\mathcal{T}(j\omega )\right]\vert =\frac{\sqrt{(\omega
^{2}+4)^{2}+4\omega ^{2}}}{\sqrt{(\omega ^{4}-6\omega ^{2}+4)^{2}+(-3\omega
^{3}+4\omega )^{2}}},
\end{equation*}%
\begin{equation*}
\int\limits_{0}^{\infty }\frac{\ln \left\vert \det [\mathcal{T}%
(s)]\right\vert }{\omega ^{2}}d\omega =0.1854.
\end{equation*}%
The system is Type [1 1] and from Eq. (20),
\begin{equation*}
\begin{split}
&\int\limits_{0}^{\infty }\frac{1}{\omega ^{2}}\ln \left\vert \det \left[ 
\mathcal{T}(j\omega )\right] \right\vert d\omega \\
=&\frac{\pi }{2}\left( \frac{1}{
-1.32+j1.53}+\frac{1}{-1.32-j1.53}+\frac{1}{-0.17+j0.97}\right.\\
&\left.+\frac{1}{%
-0.17-j0.97}-\frac{1}{-3.23}-\frac{1}{1.23}\right)+\frac{\pi}{1.236}=0.1854.
\end{split}
\end{equation*}
\section{Conclusions}
In this paper we have taken a direct approach to constraints on the sensitivity function. We have shown that the
fundamental constraint on the sensitivity function is purely a function of
the differences in speeds of the open-loop and closed-loop systems. This is
very satisfying from an intuitive point of view. The fundamental constraint
on the complementary sensitivity is a function of the closed-loop poles and
transmission zeros. The situation is made more difficult if there are RHP
poles/zeros. We hope that this Paper has contributed to understanding the
fundamental limitations in control engineering. The results have been extended
to discrete-time LTI systems.\\
\textbf{Appendix A: Proof of Theorem 1.}\\
\begin{proof}
\vspace{-0.1cm}
If the loop gain is denoted by $L(s)=\frac{N(s)}{D(s)}$, the sensitivity
function is%
\begin{equation}
\mathcal{S}(s)=\frac{1}{1+L(s)},
\end{equation}%
\begin{equation}
\begin{split}
\int\limits_{0}^{\infty }&\ln \left\vert \mathcal{S}(j\omega )\right\vert
d\omega =\int\limits_{0}^{\infty }\ln \left\vert \frac{1}{1+L(j\omega )}%
\right\vert d\omega , \\
&= \int\limits_{0}^{\infty }\ln \left\vert 1\right\vert d\omega
-\int\limits_{0}^{\infty }\ln \left\vert 1+L(j\omega )\right\vert d\omega ,
 \\
&=-\int\limits_{0}^{\infty }\ln \left\vert 1+\frac{N(j\omega )}{D(j\omega )%
}\right\vert d\omega ,   \\
&=-\int\limits_{0}^{\infty }\ln \left\vert \frac{D(j\omega )+N(j\omega )}{%
D(j\omega )}\right\vert d\omega ,   \\
&=\int\limits_{0}^{\infty }\ln \frac{\left\vert D(j\omega )\right\vert }{%
\left\vert D(j\omega )+N(j\omega )\right\vert }d\omega ,   \\
=\int\limits_{0}^{\infty }&\ln \left\vert \mathcal{S}(j\omega )\right\vert
d\omega =\int\limits_{0}^{\infty }\ln \frac{\prod\limits _{i=1}^{i=n}\left\vert (j\omega
-p_{i})\right\vert }{\prod\limits _{i=1}^{i=n}\left\vert (j\omega -\tilde{p}%
_{i})\right\vert }d\omega , \\
&=\frac{1}{2}\int\limits_{0}^{\infty }\ln \frac{\prod\limits _{i=1}^{i=n}\left\vert  (\mathrm{Re}%
\{p_{i}\}^{2}+(\omega -\mathrm{Im}\{p_{i}\})^{2})\right\vert }{\prod\limits _{i=1}^{i=n}\left\vert(\mathrm{Re}\{\tilde{p}%
_{i}\}^{2}+(\omega -\mathrm{Im}\{\tilde{p}_{i}\})^{2})\right\vert }d\omega.\\
\end{split}
\end{equation}
\vspace{-.1cm}
\begin{equation}
\begin{split}
\int & \left(\ln \frac{(x-a)^{2}+b^{2}}{(x-c)^2+d^2}+\ln \frac{(x+a)^{2}+b^{2}}{(x+c)^2+d^2}\right)dx\\
& =\ln \left(\frac{(x+a)^{2}+b^{2}}{(x-a)^2+b^2}\right)^{a}+\ln \left(\frac{(x-a)^{2}+b^{2}}{(x+c)^2+d^2}\right)^{c}\\&-(c-x)\ln \left(\frac{(a-x)^{2}+b^{2}}{(c-x)^2+d^2}\right)+x\ln \left(\frac{(x+a)^{2}+b^{2}}{(x+c)^2+d^2}\right)\\&-2b\tan^{-1}\left(\frac{a-x}{b}\right)+2b\tan^{-1}\left(\frac{a+x}{b}\right)\\
&+2d\tan^{-1}\left(\frac{c-x}{d}\right)-2d\tan^{-1}\left(\frac{c+x}{d}\right),\\
\end{split}
\end{equation}%
Let us first assume that the open-loop system is stable. Using the above integral identity, Eq. (23), we have
\begin{equation}
\begin{split}
& \int\limits_{0}^{\infty } \ln \left\vert \mathcal{S}(j\omega )\right\vert
d\omega=\frac{1}{2}\left[ -2\sum\limits _{i=1}^{i=n}\mathrm{Re}\{p_{i}\}\frac{\pi }{2}%
+2\sum\limits _{i=1}^{i=n}\mathrm{Re}\{\tilde{p}_{i}\}\frac{\pi }{2}\right] ,
 \\
&=\frac{\pi }{2}\left( \sum\limits _{i=1}^{i=n}(\tilde{p}_{i}-p_{i})\right) . 
\end{split}
\end{equation}%
Now
assume that some of the open-loop poles are in the right-hand-plane (RHP),
say $\mathrm{Re}\{p_{i}\}>0,i=1,2,...,n_{p}$. There will be additional nonzero terms and we have that%
\begin{equation}
\begin{split}
&\int\limits_{0}^{\infty }\ln \left\vert \mathcal{S}(j\omega )\right\vert
d\omega\\
& = \frac{1}{2}\left[ -2\sum\limits _{i=1}^{i=n}\mathrm{Re}\{p_{i}\}\frac{\pi 
}{2}+2\sum\limits _{i=1}^{i=n}\mathrm{Re}\{\tilde{p}_{i}\}\frac{\pi }{2}+\sum\limits
_{i=1}^{i=n_{p}}2\mathrm{Re}\{p_{i}\}\pi \right] , \\
&=\frac{\pi }{2}\left( \sum\limits _{i=1}^{i=n}(\mathrm{Re}\{\tilde{p}_{i}\}-\mathrm{Re}\{p_{i}\})+\sum\limits _{i=1}^{i=n_{p}}2\mathrm{Re}\{p_{i}\}\right) . 
\end{split}
\end{equation}%
\end{proof}
\textbf{Appendix B: Proof of Theorem 2.}\\
\begin{proof}
The complementary sensitivity
function may be written as%
\begin{equation}
\mathcal{T}(s)=K\frac{(s-z_{1})(s-z_{2})...(s-z_{m})}{(s-\tilde{p}_{1})(s-%
\tilde{p}_{2})...(s-\tilde{p}_{n})},\qquad m\leq n
\end{equation}%
\begin{equation}
\begin{split}
&\int\limits_{0}^{\infty }\frac{\ln \left\vert \mathcal{T}(j\omega
)\right\vert }{\omega ^{2}}d\omega = \int\limits_{0}^{\infty }\frac{\ln
\left\vert K\frac{z_{1}z_{2}...z_{m}}{\tilde{p}_{1}\tilde{p}_{2}...\tilde{p}%
_{n}}\frac{\prod\limits _{i=1}^{i=m}(j\frac{\omega }{\mathrm{Re}\{z_{i}\}+j\mathrm{Im}%
\{z_{i}\}}-1)}{\prod\limits _{i=1}^{i=n}(j\frac{\omega }{\mathrm{Re}\{\tilde{p}_{i}\}+j%
\mathrm{Im}\{\tilde{p}_{i}\}}-1)}\right\vert }{\omega ^{2}}d\omega , \\
&=\int\limits_{0}^{\infty }\ln\frac{\left\vert K\frac{z_{1}z_{2}...z_{m}}{%
\tilde{p}_{1}\tilde{p}_{2}...\tilde{p}_{n}}\right\vert }{\omega ^{2}}d\omega\\
&+\int\limits_{0}^{\infty }\frac{\ln\frac{\prod\limits _{i=1}^{i=m}\left\vert\tilde{p}_{i}\right\vert \left[\left(\omega-\mathrm{Im}(z_i) \right)
^{2}+\left(\mathrm{Re}(z_{i})\right)^2\right]^{\frac{1}{2}}}{\prod\limits _{i=1}^{i=n}\left\vert
z_{i}\right\vert\left[ \left(\omega-\mathrm{Im}(\tilde p_{i})\right)
^{2}+\left(\mathrm{Re}(\tilde{p}_{i})\right) ^{2}\right] ^{\frac{1}{2}}}}{\omega^2}d\omega ,  \\
&=\int\limits_{0}^{\infty }\ln\frac{\left\vert K\frac{z_{1}z_{2}...z_{m}}{%
\tilde{p}_{1}\tilde{p}_{2}...\tilde{p}_{n}}\right\vert }{\omega ^{2}}d\omega\\
&+\frac{1}{2}\int\limits_{0}^{\infty }\frac{\ln\frac{\prod\limits _{i=1}^{i=m}\left\vert
\tilde{p}_{i}\right\vert^2\left[ \left( \omega-\mathrm{Im}(z_i) \right)
^{2}+\left( \mathrm{Re}(z_{i})\right)^2\right]}{\prod\limits _{i=1}^{i=n}\left\vert
z_{i}\right\vert^2\left[ \left(\omega-\mathrm{Im}(\tilde p_{i})\right)
^{2}+\left(\mathrm{Re}(\tilde{p}_{i})\right) ^{2}\right] }}{\omega^2}d\omega ,  \\
\end{split}
\end{equation}%
If the system is Type 0, then the first term in the integral becomes
unbounded,
\begin{equation}
\int\limits_{0}^{\infty }\frac{\ln \left\vert \mathcal{T}(j\omega
)\right\vert }{\omega ^{2}}d\omega =\infty .
\end{equation}%
If the system is Type I (or higher), the first term is zero. Let us first assume that the system is minimum-phase. Using the integral identity
\begin{equation}
\begin{split}
&\int \frac{\ln (\frac{(c^2+d^2)^{2}((-x^{2}+a^{2}+b^{2})^2+4a^{2}x^{2}) }{ (a^2+b^2)^{2}((-x^{2}+c^{2}+d^{2})^2+4c^{2}x^{2}) }}
{x^{2}}dx=\\
&\frac{\ln (\frac{(c^2+d^2)^{2}((-x^{2}+a^{2}+b^{2})^2+4a^{2}x^{2}) }{ (a^2+b^2)^{2}((-x^{2}+c^{2}+d^{2})^2+4c^{2}x^{2}) }}
{x}\\
&+\frac{2\tan ^{-1}\left( \frac{x}{a-jb}\right) }{a-jb}+\frac{2\tan ^{-1}\left( \frac{x}{a+jb}\right) }{a+jb}\\&-\frac{2\tan ^{-1}\left( \frac{x}{c-jd}\right) }{c-jd}-\frac{2\tan ^{-1}\left( \frac{x}{c+jd}\right) }{c+jd},
\end{split}
\end{equation}
For a Type I (or higher) system we then obtain
\begin{equation}
\begin{split}
&\int\limits_{0}^{\infty }\frac{\ln \left\vert \mathcal{T}(j\omega
)\right\vert }{\omega ^{2}}d\omega \\
&=-\frac{\pi }{2}\sum\limits\limits_{i=1}^{i=m}\frac{%
1}{2}\frac{2\mathrm{Re}(z_{i})}{(\mathrm{Re}(z_{i}))^{2}+(\mathrm{Im}(z_{i}))^{2}}+%
\frac{\pi }{2}\sum\limits\limits_{i=1}^{i=n}\frac{1}{2}\frac{2\mathrm{Re}(\tilde{p}_{i})}{%
(\mathrm{Re}(\tilde{p}_{i}))^{2}+\left(\mathrm{Im}(\tilde{p}_{i})\right)^{2}},\\
&=\frac{\pi }{2}\left( \sum\limits_{i=1}^{i=m}-\frac{1}{z_{i}}+\sum\limits_{i=1}^{i=n}%
\frac{1}{\tilde{p}_{i}}\right). 
\end{split}
\end{equation}
If the system has non-minimum phase zeros, there are additional non-zero terms. Again for a Type I (or higher) system we obtain
\begin{equation}
\begin{split}
&\int\limits_{0}^{\infty }\frac{\ln \left\vert \mathcal{T}(j\omega
)\right\vert }{\omega ^{2}}d\omega =-\frac{\pi }{2}\sum\limits\limits_{i=1}^{i=m}\frac{%
1}{2}\frac{2\mathrm{Re}(z_{i})}{(\mathrm{Re}(z_{i}))^{2}+(\mathrm{Im}(z_{i}))^{2}}\\
&+\frac{\pi }{2}\sum\limits\limits_{i=1}^{i=n}\frac{1}{2}\frac{2\mathrm{Re}(\tilde{p}_{i})}{%
(\mathrm{Re}(\tilde{p}_{i}))^{2}+(\mathrm{Im}(\tilde{p}_{i}))^{2}}\\
&+\pi \sum\limits_{i=1}^{i=n_{z}}\frac{1}{2}\frac{2\mathrm{Re}(z_{i})}{(\mathrm{Re}%
(z_{i}))^{2}+(\mathrm{Im}(z_{i}))^{2}},  \\
&=\frac{\pi }{2}\left( \sum\limits_{i=1}^{i=m}-\frac{1}{z_{i}}+\sum\limits_{i=1}^{i=n}%
\frac{1}{\tilde{p}_{i}}\right) +\pi \sum\limits_{i=1}^{i=n_{z}}\frac{1}{z_{i}}. 
\end{split}
\end{equation}
\end{proof}
\textbf{Appendix C: Proof of Theorem 3.}\\
\begin{proof}
\begin{equation}
\begin{split}
&\int\limits_{0}^{\infty }\ln \left\vert \det \left[ \mathcal{S}(j\omega )%
\right] \right\vert d\omega =\int\limits_{0}^{\infty }\ln \left\vert \det %
\left[ I+L(j\omega )\right] ^{-1}\right\vert d\omega\\
&\int\limits_{0}^{\infty }\ln \left\vert \frac{1}{\det \left[ I+L(j\omega )%
\right] }\right\vert d\omega
=-\int\limits_{0}^{\infty }\ln \left\vert \det \left[ I+L(j\omega )\right]
\right\vert d\omega . 
\end{split}
\end{equation}%
Suppose the loop gain $L(s)$ is written as an irreducible right-matrix
fraction description (MFD) [31, page 447] 
\begin{equation}
L(s)=N(s)D(s)^{-1},
\end{equation}
\begin{equation}
\begin{split}
&\int\limits_{0}^{\infty }\ln \left\vert \det \left[ \mathcal{S}(j\omega )%
\right] \right\vert d\omega \\
&=-\int\limits_{0}^{\infty }\ln \left\vert
\det \left[ I+N(j\omega )D(j\omega )^{-1}\right] \right\vert d\omega , \\
&=-\int\limits_{0}^{\infty }\ln \left\vert \det \left[ D(j\omega
)+N(j\omega )\right] \right\vert \det \left[ D(j\omega )^{-1}\right] d\omega
,  \\
&=-\int\limits_{0}^{\infty }\ln \left\vert \frac{\det \left[ D(j\omega
)+N(j\omega )\right] }{\det \left[ D(j\omega )\right] }\right\vert d\omega ,
 \\
&=-\int\limits_{0}^{\infty }\ln \left\vert \frac{\det \left[ D(j\omega
)+N(j\omega )\right] }{\det \left[ D(j\omega )\right] }\right\vert d\omega ,
 \\
&=-\int\limits_{0}^{\infty }\ln \left\vert \frac{\phi _{cl}(j\omega )}{%
\phi _{ol}((j\omega )}\right\vert d\omega ,  
\end{split}
\end{equation}%
where $\phi _{ol}(s)$ and $\phi _{cl}(s)$\ are the open-loop and closed-loop
characteristic polynomials of the system.%
\begin{equation}
\begin{split}
\int\limits_{0}^{\infty }&\ln \left\vert \det \left[ \mathcal{S}(j\omega )%
\right] \right\vert d\omega=\int\limits_{0}^{\infty }\ln \frac{\prod\limits _{i=1}^{i=n}\left\vert (j\omega
-p_{i})\right\vert }{\prod\limits _{i=1}^{i=n}\left\vert (j\omega -\tilde{p}%
_{i})\right\vert }d\omega , \\
&=\frac{1}{2}\int\limits_{0}^{\infty }\ln \frac{\prod\limits _{i=1}^{i=n}\left\vert  (\mathrm{Re}%
\{p_{i}\}^{2}+(\omega -\mathrm{Im}\{p_{i}\})^{2})\right\vert }{\prod\limits _{i=1}^{i=n}\left\vert(\mathrm{Re}\{\tilde{p}%
_{i}\}^{2}+(\omega -\mathrm{Im}\{\tilde{p}_{i}\})^{2})\right\vert }d\omega . \\
\end{split}
\end{equation}%
Let us assume for the moment that the open-loop system is stable. Using the integral identity (23) we have
\begin{equation}
\begin{split}
& \int\limits_{0}^{\infty } \ln  \left\vert \det \left[ \mathcal{S}(j\omega )%
\right] \right\vert
d\omega\\
&=\frac{1}{2}\left[ -2\sum\limits _{i=1}^{i=n}\mathrm{Re}\{p_{i}\}\frac{\pi }{2}%
+2\sum\limits _{i=1}^{i=n}\mathrm{Re}\{\tilde{p}_{i}\}\frac{\pi }{2}\right] ,
 \\
&=\frac{\pi }{2}\left( \sum\limits _{i=1}^{i=n}(\tilde{p}_{i}-p_{i})\right) . 
\end{split}
\end{equation}%
Now
assume that some of the open-loop poles are in the right-hand-plane (RHP),
say $\mathrm{Re}\{p_{i}\}>0,i=1,2,...,n_{p}$. Then we have that%
\begin{equation}
\begin{split}
&\int\limits_{0}^{\infty }\ln  \left\vert \det \left[ \mathcal{S}(j\omega )%
\right] \right\vert
d\omega\\
& = \frac{1}{2}\left[- 2\sum\limits _{i=1}^{i=n}\mathrm{Re}\{p_{i}\}\frac{\pi 
}{2}+2\sum\limits _{i=1}^{i=n}\mathrm{Re}\{\tilde{p}_{i}\}\frac{\pi }{2}+2\sum\limits
_{i=1}^{i=n_{p}}\mathrm{Re}\{p_{i}\}\pi \right] , \\
&=\frac{\pi }{2}\left( \sum\limits _{i=1}^{i=n}(\mathrm{Re}\{\tilde{p}_{i}\}-\mathrm{Re}\{p_{i}\})+\sum\limits _{i=1}^{i=n_{p}}2\mathrm{Re}\{p_{i}\}\right) . 
\end{split}
\end{equation}%
\end{proof}
\textbf{Appendix D: Proof of Theorem 4.}\\
\begin{proof}
Suppose the loop gain $L(s)$ is written as an irreducible right-matrix
fraction description (MFD) [30, page 447] 
\begin{equation}
L(s)=N(s)D(s)^{-1},
\end{equation}
then
\begin{equation}
\begin{split}
&\int\limits_{0}^{\infty }\frac{1}{\omega ^{2}}\ln \left\vert \det \left[ 
\mathcal{T}(j\omega )\right] \right\vert d\omega \\
&=\int\limits_{0}^{\infty
}\frac{1}{\omega ^{2}}\ln \left\vert \frac{\det [N(j\omega )]}{\det
[D(j\omega )+N(j\omega )]}\right\vert d\omega , \\
&=\int\limits_{0}^{\infty }\frac{1}{\omega ^{2}}\ln \left\vert \frac{\phi
_{z}(j\omega )}{\phi _{cl}(j\omega )}\right\vert d\omega ,  
\end{split}
\end{equation}
where the zeros of the $\phi _{z}(s)$ polynomial are the transmission zeros
of the system and $\phi _{cl}(s)$ is the closed-loop characteristic
polynomial.
\begin{equation}
\begin{split}
&\int\limits_{0}^{\infty }\frac{\ln \left\vert \det \left[ \mathcal{T}%
(j\omega )\right] \right\vert }{\omega ^{2}}d\omega\\
&= \int\limits_{0}^{\infty }\frac{\ln
\left\vert K\frac{z_{1}z_{2}...z_{m}}{\tilde{p}_{1}\tilde{p}_{2}...\tilde{p}%
_{n}}\frac{\prod\limits _{i=1}^{i=m}(j\frac{\omega }{\mathrm{Re}\{z_{i}\}+j\mathrm{Im}%
\{z_{i}\}}-1)}{\prod\limits _{i=1}^{i=n}(j\frac{\omega }{\mathrm{Re}\{\tilde{p}_{i}\}+j%
\mathrm{Im}\{\tilde{p}_{i}\}}-1)}\right\vert }{\omega ^{2}}d\omega , \\
\end{split}
\end{equation}
\begin{equation}
\begin{split}
&=\int\limits_{0}^{\infty }\ln\frac{\left\vert K\frac{z_{1}z_{2}...z_{m}}{%
\tilde{p}_{1}\tilde{p}_{2}...\tilde{p}_{n}}\right\vert }{\omega ^{2}}d\omega\\
&+\int\limits_{0}^{\infty }\frac{\ln\frac{\prod\limits _{i=1}^{i=m}\left\vert\tilde{p}_{i}\right\vert \left[\left(\omega-\mathrm{Im}(z_i) \right)
^{2}+\left(\mathrm{Re}(z_{i})\right)^2\right]^{\frac{1}{2}}}{\prod\limits _{i=1}^{i=n}\left\vert
z_{i}\right\vert\left[ \left(\omega-\mathrm{Im}(\tilde p_{i})\right)
^{2}+\left(\mathrm{Re}(\tilde{p}_{i})\right) ^{2}\right] ^{\frac{1}{2}}}}{\omega^2}d\omega ,  \\
\end{split}
\end{equation}
\begin{equation}
\begin{split}
&=\int\limits_{0}^{\infty }\ln\frac{\left\vert K\frac{z_{1}z_{2}...z_{m}}{%
\tilde{p}_{1}\tilde{p}_{2}...\tilde{p}_{n}}\right\vert }{\omega ^{2}}d\omega\\
&+\frac{1}{2}\int\limits_{0}^{\infty }\frac{\ln\frac{\prod\limits _{i=1}^{i=m}\left\vert
\tilde{p}_{i}\right\vert^2\left[ \left( \omega-\mathrm{Im}(z_i) \right)
^{2}+\left( \mathrm{Re}(z_{i})\right)^2\right]}{\prod\limits _{i=1}^{i=n}\left\vert
z_{i}\right\vert^2\left[ \left(\omega-\mathrm{Im}(\tilde p_{i})\right)
^{2}+\left(\mathrm{Re}(\tilde{p}_{i})\right) ^{2}\right] }}{\omega^2}d\omega.  \\
\end{split}
\end{equation}%
If the system is Type [0 0\ldots 0] then the first term in the integral
becomes unbounded. 
\begin{equation}
\int\limits_{0}^{\infty }\frac{\ln \left\vert \det \left[ \mathcal{T}%
(j\omega )\right] \right\vert }{\omega ^{2}}d\omega =\infty.
\end{equation}%
For a Type [1 1\ldots1] (or higher) system 
\begin{equation}
\left\vert \frac{Kz_{1}z_{2}\ldots z_{m}}{\tilde{p}_{1}\tilde{p}_{2}\ldots\tilde{p}%
_{n}}\right\vert =1.
\end{equation}%
Using the integral identity (29) for a Type [1 1 \ldots 1] (or higher) system we obtain%
\begin{equation}
\begin{split}
&\int\limits_{0}^{\infty }\frac{\ln \left\vert \det \left[ \mathcal{T}%
(j\omega )\right] \right\vert }{\omega ^{2}}d\omega=
-\frac{\pi }{2}%
\sum\limits_{i=1}^{i=m}\frac{1}{2}\frac{2\mathrm{Re}(z_{i})}{(\mathrm{Re}(z_{i}))^{2}+%
(\mathrm{Im}(z_{i}))^{2}}\\
&+\frac{\pi }{2}\sum\limits_{i=1}^{i=n}\frac{1}{2}\frac{2\mathrm{%
Re}(\tilde{p}_{i})}{(\mathrm{Re}(\tilde{p}_{i}))^{2}+(\mathrm{Im}(\tilde{p}_{i}))^{2}%
}\\
&+\pi \sum\limits_{i=1}^{i=n_{z}}\frac{1}{2}\frac{2\mathrm{Re}(z_{i})}{(\mathrm{Re}%
(z_{i}))^{2}+(\mathrm{Im}(z_{i}))^{2}},   \\
&=\frac{\pi }{2}\left( -\sum\limits_{i=1}^{i=m}\frac{1}{z_{i}}+\sum\limits_{i=1}^{i=n}%
\frac{1}{\tilde{p}_{i}}\right) +\pi \sum\limits_{i=1}^{i=n_{z}}\frac{1}{z_{i}}. 
\end{split}
\end{equation}
\end{proof}
\vspace{-0.2cm}
\section*{Acknowledgments}
The authors gratefully acknowledge the help of Drs. S. A. McCabe and Jun-Kyu Lee of SC Solutions.


\bibliographystyle{IEEEtran}

\end{document}